\documentclass[twocolumn,prl,showpacs]{revtex4}
\usepackage{graphicx}
\begin{document}


\title{Resonantly Enhanced Axion-Photon Regeneration}

\author{P. Sikivie,$^{a,b}$ D.B. Tanner,$^a$ and Karl van Bibber$^c$}

\affiliation{
$^a$~Department of Physics, University of Florida, 
Gainesville, FL 32611, USA\\
$^b$~Theoretical Physics Division, CERN, CH-1211 Gen\`eve 23, 
Switzerland\\
$^c$~Lawrence Livermore National Laboratory,
Livermore, CA 94550, USA
}

\begin{abstract}

We point out that photon regeneration-experiments that search for the
axion, or axion-like particles, may be resonantly enhanced by employing
matched Fabry-Perot optical cavities encompassing both the axion production 
and conversion magnetic field regions.   Compared to a simple photon
regeneration experiment, which uses the laser in a
single-pass geometry, this technique can result in a gain in rate of
order ${\cal F}^2$, where ${\cal F}$  is the finesse of the cavities.  
This gain could feasibly be $10^{(10-12)}$, corresponding to an 
improvement in sensitivity in the axion-photon coupling, 
$g_{a\gamma\gamma}$ , of order ${\cal F}^{1/2} \sim 10^{(2.5-3)}$, permitting a
practical purely laboratory search to probe axion-photon couplings not
previously excluded by stellar evolution limits, or solar axion searches.

\end{abstract}

\maketitle

The 
axion remains the most attractive solution to the strong-CP
problem and is one of two leading dark-matter candidates \cite{Brad03}.  
Recently, it has been realized that the axion represents a fundamental
underlying feature of string theories; there could be several or even a
great number of axions or axion-like particles within any particular
string theory \cite{Svrc06}.  From the experimental viewpoint, there 
are now several photon regeneration experiments in various stages of
preparation \cite{Raba06}, motivated in part by the report of the PVLAS
collaboration of a non-zero magnetically-induced dichroism 
of the vacuum \cite{Zava06}, which may be interpreted as the 
production of a light boson with a two-photon coupling \cite{Siki83}.  
Although the particle interpretation of the PVLAS experiment 
is in principle 
excluded by the much more stringent limit set by the CAST solar 
axion search \cite{Ziou05}, it is important to check this 
result by a purely laboratory experiment, particularly if 
one could ultimately improve such a measurement to reach previously
unexplored regions of $(m_a,g_{a\gamma\gamma})$.

The simplest 
and most unambiguous purely laboratory experiment to look for light
pseudoscalars is 
photon regeneration (``shining light through walls'') 
\cite{vanB87}.  A laser beam traverses a dipole magnet,
wherein a small fraction of the photons are converted into axions with 
the same energy.  An optical barrier blocks the primary laser beam,
whereas the axion component of the beam travels through the wall 
unimpeded and enters
a second dipole
magnet, where it is reconverted to  photons with the same probability
(we will assume magnets of identical length $L$ and field strength 
$B_0$ without loss of generality).   However, because the photon-regeneration
rate goes as $g_{a\gamma\gamma}^4$,  the sensitivity of
the experiment to small values of $g_{a\gamma\gamma}$ is poor 
in its basic form, improved appreciably only by increasing $B_0$ or $L$.  

In this Letter, we
point out that matched Fabry-Perot cavities incorporated into the production
and detection magnets can improve the sensitivity in the axion-photon
coupling, $g_{a\gamma\gamma}$ by the square root 
of  the cavities' finesse, ${\cal F}^{1/2}$.
This factor may be $10^{(2.5-3)}$. 

The action density for the dynamics of photons and axions is
\begin{equation}
{\cal L} = {1 \over 2} (\epsilon E^2 - B^2)
+ {1 \over 2} (\partial_t a)^2 - {1 \over 2} (\vec{\nabla} a)^2
- {1 \over 2} m_a^2 a^2 - g a \vec{E} \cdot \vec{B}
\label{actden}
\end{equation}
where $\vec{E}$, $\vec{B}$ and $a$ are respectively the
electric, magnetic and axion fields.  The electromagnetic
fields are given in terms of scalar and vector potentials,
$\vec{E} = - \vec{\nabla} \Phi - \partial_t \vec{A},~
\vec{B} = \vec{\nabla} \times \vec{A}$, as usual.  The 
coupling $g \equiv g_{a\gamma\gamma}$ is written without 
subscripts here, for simplicity.  $\epsilon$ is assumed 
constant in both space and time.  In the presence of a 
large static magnetic field $\vec{B}_0(\vec{x})$, the 
equations of motion are
\begin{eqnarray}
\epsilon \vec{\nabla} \cdot \vec{E} &=&
g \vec{B}_0 \cdot \vec{\nabla} a \nonumber\\
\vec{\nabla} \times \vec{B} - \epsilon \partial_t \vec{E} &=&
- g \vec{B}_0 \partial_t a \nonumber\\
\partial_t^2 a - \vec{\nabla}^2 a + m_a^2 a &=&
- g  \vec{E} \cdot \vec{B_0}~~~~\ .
\label{eom}
\end{eqnarray}
$\vec{B}$ now represents the magnetic field minus $\vec{B}_0$,
and terms of order $gE$ and $gB$ are neglected.  Eqs.~(\ref{eom})
describe the conversion of axions to photons and vice-versa.

Using these equations, it can be shown \cite{Siki83,vanB87,Raff88} 
that the axion-to-photon conversion probability in a region of 
length $L$, permeated by a constant magnetic field $B_0$ transverse 
to the direction of propagation and a dielectric constant $\epsilon$,
is given by ($\hbar = c = 1$)
\begin{equation}
p = {1 \over 4 \beta_a \sqrt{\epsilon}}(g B_0 L)^2
\left({2 \over q L} \sin {q L \over 2}\right)^2~~~~\ ,
\label{prob}
\end{equation}
with $\beta_a$ the axion speed and $q = k_a - k_\gamma$
the momentum transfer.  In terms of the energy $\omega$,
which is the same for the axion and the photon,
$k_a = \sqrt{\omega^2 - m_a^2}$, $\beta_a = {k_a \over \omega}$ 
and $k_\gamma = \sqrt{\epsilon} \omega$.  The photon to axion 
conversion probability in this same region is also equal to $p$.  
Everything else being the same, the conversion probability is 
largest when $q \approx 0$.  For $m_a << \omega$, and propagation in a vacuum
\begin{equation}
q = - {m_a^2 \over 2 \omega} + (1 - \sqrt{\epsilon}) \omega~~~\ .
\label{q}
\end{equation}

Fig. 1a shows the axion-photon regeneration 
experiment as usually conceived.
If $P_0$ is the power of the laser, the power of the axion beam 
traversing the wall is $p~P_0$ where $p$ is the conversion
probability in the magnet on the LHS of Fig. 1a.  Let $p^\prime$ 
be the conversion probability in the magnet on the RHS.  The power 
in regenerated photons is $P = p^\prime~p~P_0$.

Fig. 1b shows the two improvements we propose for the experiment.  
The first improvement is to build up the power 
on the photon-to-axion 
conversion side of the experiment 
using a Fabry-Perot cavity, as illustrated.  Photons 
in the production cavity will then convert to axions with probability $p$ 
for each pass through the cavity.  The standing wave in the production 
cavity is the sum of left-moving and right-moving components of 
equal amplitude.  If the reflectivity of the cavity mirrors is given by
\begin{equation}
R = 1 - \eta
\label{refl}
\end{equation}
and the power of the laser is $P_0$, the power of the 
right-moving wave in the production cavity is ${1 \over \eta}~P_0$.
Therefore the axion power through the wall in the setup
of Fig. 1b is ${1 \over \eta}~p~P_0$.  Assuming the lasers 
in Fig. 1a and Fig. 1b have the same power, the axion flux 
is increased by the factor ${1 \over \eta}$.

\begin{figure}
\includegraphics[width=0.9\columnwidth]{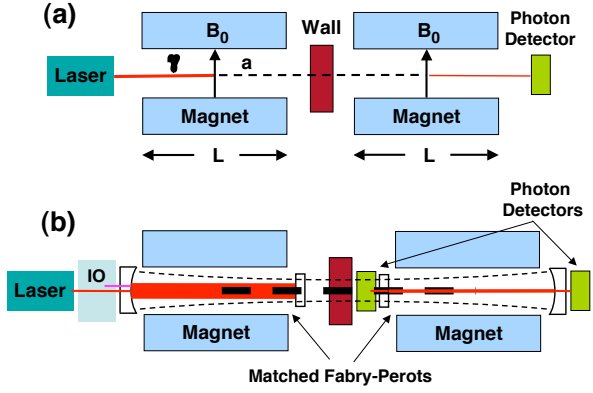}
\caption{(color) (a)  Simple photon regeneration.   (b)  Resonant
photon regeneration, employing matched Fabry-Perot cavities.  The
overall envelope schematically shown by the thin dashed lines
indicates the important condition that the axion wave, and thus the
Fabry-Perot mode, in the conversion magnet must follow that of the
hypothetically unimpeded photon wave from the Fabry-Perot mode in
the production magnet. Between the laser and the cavity is the 
injection optics (IO) which manages mode matching of the laser to the 
cavity, imposes RF sidebands for reflection locking of the laser to 
the cavity, and provides isolation for the laser.
The photon detectors are also preceded by matching and beam-steering 
optics. 
Not shown at all is the electro-optical system required to lock the two
cavities together in frequency.}
\end{figure}

Increasing the axion production rate, and thus the photon regeneration
rate, by building up the optical power in the first magnet is not a new
idea.  In fact, the only photon regeneration experiment performed and published to date,
by Ruoso et al. utilized an ``optical delay line,'' i.e., an incoherent
cavity encompassing the production magnet, causing the laser beam to
traverse the magnet  200 times before exiting.  With relatively modest
magnets (4.4 m, 3.7 T each), a limit of   
$g_{a\gamma\gamma} < 7.7 \times 10^{-7}~{\rm GeV}^{-1}$ 
was set \cite{Ruos92}.

There is substantial gain from
building up the laser
power in the axion production magnet; however, it is immaterial whether one
``recycles'' the photons incoherently, as in an optical delay line, or
coherently, as in a Fabry-Perot (FP) cavity. In contrast, the coherent case 
alone can provide a large
additional gain in sensitivity
for photon regeneration.  
Thus, the second improvement is to also install a Fabry-Perot cavity on
the regeneration side of the experiment, making a symmetric arrangement,
as illustrated in Fig. 1b.   When the second FP cavity is
locked to the first, the probability of axion 
to photon conversion in the second FP cavity is 
${2 \over {\eta}^\prime}~p^\prime  = 
{2 \over \pi}~{\cal F}^\prime ~p^\prime$ 
where ${\cal F}^\prime$ is the finesse 
of the cavity, and $p^\prime$ is the axion-to-photon conversion
probability in the absence of the cavity.  The calculation which yields
this result is outlined in the next paragraph.

The cavity modes are described by
\begin{equation}
\vec{A}_n = A_n(t) \hat{y} \sin({n \pi \over L} z)
\label{FPA}
\end{equation}
where $\hat{z}$ is in the direction of light propagation and
$\hat{y}$ is a transverse direction.  The dependence of the
mode function on the transverse coordinates ($x$ and $y$) is
neglected here, but will be discussed later.  Using Eqs~(\ref{eom})
one can show that, in the presence of an axion beam travelling through
the cavity in the $z$-direction
\begin{equation}
a(z,t) = {\cal A} \sin (k_a z - \omega t)~~~~\ ,
\label{axbeam}
\end{equation}
the coefficients $A_n(t)$ satisfy
\begin{equation}
({d^2 \over dt^2} + \gamma {d \over dt} + \omega_n^2) A_n(t)
= C \sin(\omega t - {qL \over 2})~~~~\ ,
\label{eom7}
\end{equation}
where $\omega_n = {n \pi \over \sqrt{\epsilon} L}$ and
\begin{equation}
C = {1 \over \epsilon} g \omega B_0 {\cal A}
{2 \over Lq} \sin({qL \over 2})~~~\ .
\label{Fn}
\end{equation}
As before, $q = k_a - k_n = \sqrt{\omega^2 - m_a^2} - {n \pi \over L}$
is the momentum transfer.  When the production cavity and the regeneration
cavity are tuned to the same frequency, $\omega_n = \omega$ for some
$n$.  Then
\begin{equation}
\vec{A}_n = \hat{y}  {C \over \omega \gamma}
\sin({n \pi \over L} z) 
\sin(\omega t - {q L \over 2} - {\pi \over 2})~~\ ,
\label{sol}
\end{equation}
up to transients.  The energy stored in the cavity is
$E = {1 \over 4} S L A^2 \epsilon \omega^2$ where
$A = {C \over \omega \gamma}$ and $S$ is the cross-sectional area
of the cavity mode.  The power emitted by the cavity is $P = \gamma E$,
assuming that there are no losses other than by transmission through
the mirrors.  The power of the axion beam is
$P_a = {1 \over 2} S {\cal A}^2 \omega k_a$, assuming it has the
same cross-sectional area $S$ as the cavity mode.  The axion to 
photon conversion probability in the FP cavity is therefore
\begin{equation}
p_{\rm FP} = {P \over P_a} = {1 \over 2}
{g^2 B_0^2 \over \epsilon \beta_a \omega}
Q L \left({2 \over qL} \sin({qL \over 2})\right)^2~~~~\ ,
\label{FPprob}
\end{equation}
where $Q = {\omega \over \gamma}$ is the quality factor of the cavity.
In terms of the conversion probability $p^\prime$ in the same region
(length $L$, magnetic field $B_0$, and dielectric constant $\epsilon$)
without the cavity, we have
$p_{\rm FP} = {2 Q \over \sqrt{\epsilon} L \omega} p^\prime =
{2 {\cal F}^\prime \over \pi} p^\prime$ as announced.

Including both improvements, the regenerated photon power emitted 
from the cavity is 
\begin{equation}
P = {2 \over \eta^\prime \eta}~p^\prime~p~P_0~~~\ .
\label{regpow}
\end{equation}
Half of the power $P$ is right-moving and half is left-moving.
To detect all the regenerated photons, detectors are installed 
on both sides of the Fabry-Perot cavity, as illustrated.  The 
combined improvements yield an increase by a factor 
${2 /\eta \eta^\prime}$ in signal power.  With present 
technology, this factor may be as large as $10^{12}$.

In general, the loss of power from the cavity will have other
contributions, $\eta^\prime  =  \eta^\prime_{\rm trans} + 
\eta^\prime_{abs} + \eta^\prime_{scatt}$, where the latter
two terms represent absorption and scattering (including 
diffraction) losses.  If 
transmission is not entirely dominant, 
then Eq. (\ref{regpow}) should be modified by a multiplicative 
factor $f  = \eta^\prime_{\rm trans}/\eta^\prime$.    There 
is no such factor corresponding to the production side, as the 
transmitted power is irrelevant, and the laser power 
can always be brought up to the limit established by optical
damage.

There are practical limitations to the optics which  establish the
achievable sensitivity of a resonant photon regeneration experiment. 
Here, we present a plausible experimental realization, utilizing 
a 10~W CW Nd:YAG ($\lambda = 1.064$~$\mu$m) laser, similar to the 
LIGO laser \cite{LIGO-NIM}, and a total of eight LHC dipole magnets 
(8.75 T, 14.3 m, 50 mm $\oslash$), which are well-suited to the 
experiment.

With the magnets set end-to-end, four for the production leg and
four for the regeneration leg, the overall length of each cavity 
will exceed 60 m.  We use 66 m in this estimate, giving a cavity 
free spectral range (FSR) of 2.3 MHz.  We will refer to this as a 
4 + 4 configuration of LHC dipole magnets.  The magnet diameter 
determines the maximum size of the TEM$_{00}$ Gaussian mode of the 
cavities.  Furthermore, the spatio-temporal profiles of the axion 
and photon modes are identical, {\it i.e.,} the axion mode
follows that of a hypothetically unobstructed photon beam.  Thus 
the cavities should be symmetric, with the beam waist
at the optical barrier,  such that the two end mirrors would support the
optical mode if the barrier and inboard mirrors were removed
(Figure 1b).  There is no constraint that the length of the two cavities
be exactly equal, nor that their relative separation equal a multiple
number of wavelengths.  For the case of 1.064 $\mu$m light, and for 
130~m between end mirrors, the beam is everywhere smaller than 36 mm, 
and is not clipped by the sagitta of the LHC dipoles \cite{Pugn06}.

Another limit is that the intracavity power density at 
the mirrors be below the damage threshold, which for the best
multilayer dielectric mirrors approaches 1 GW/cm$^2$.  Putting the waist 
of the mode at the barrier implies that the limitation on power density
will be first encountered for the slightly convex inboard mirrors.
The total circulating power in the production magnet is $\sim$1~MW.  

The best Fabry-Perot resonators have achieved a finesse of a few
million; here, we take for the finesse, ${\cal F} \sim 3.1 \times 10^5$,
an order of magnitude lower, which is still ambitious, but feasible.  
For the line-width implied by this finesse, 7.3 Hz, the vibration
tolerance for the cavity mirrors is of order $10^{-3}$ nm, well 
within the experience of LIGO detectors \cite{LIGO-NIM}.   Also 
for this finesse, we can design the cavity to ensure that transmission
dominates loss (scattering and diffraction), and this has been assumed.
 
Intrinsic to resonantly enhanced photon regeneration is the requirement
that the production and regeneration cavities remained locked in frequency
together, within their bandwidth, $\Delta \nu = 2\nu/Q$, where the quality
factor, $Q = n{\cal F}$ of the cavity, and $n$ its mode index. The 10~W laser is 
reflection locked to one of the cavity modes of the production cavity
using the Pound-Dreever-Hall reflection locking scheme \cite{PDH}.
Then, the challenge is to implement 
the locking of the regeneration cavity
in a way that does not introduce
any spurious photons into the detectors.  The scheme we
envision is to use a low power Nd:YAG laser, offset locked by 
(integer)$\times$FSR of the production cavity (say, 50 MHz) from 
the main laser, and use the same Pound-Dreever-Hall reflection 
locking method to control the length of the regeneration cavity. 
For additional rejection, the locking could be done in the orthogonal
polarization state, i.e. perpendicular to the dipole magnetic field
${\vec B_0}$.

There are several possible schemes for detection of the weak signal from
the regeneration cavity. The simplest is to focus the cavity output on a
cooled InGaAs charge-coupled devices (CCDs), with pixels of the order of
15 $\mu$m. Modern CCDs developed for astronomy have very low dark current
rates, of order 1-2 $e~{\rm min}^{-1} {\rm pixel}^{-1}$. However, the best
scheme seems to be heterodyne detection, mixing the generation cavity
output with the locking laser at an RF photodiode and detecting both
in-phase and quadrature signals at the difference frequency between the
laser in the generating cavity and the locking beam, especially as the
frequencies of both lasers (and their difference) are well known.

\begin{figure}
\includegraphics[width=0.94\columnwidth]{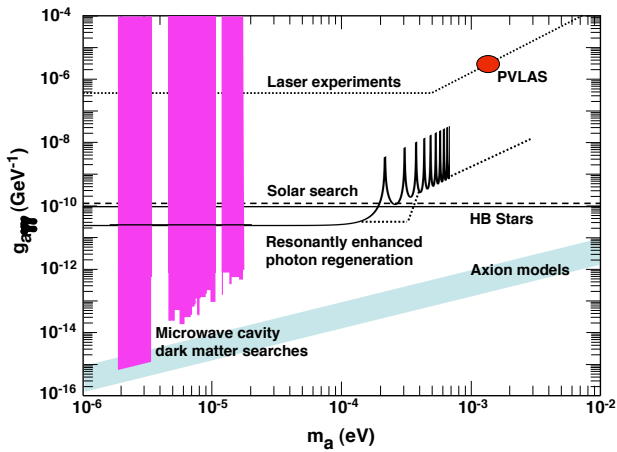}
\caption{(color) Current exclusion plot of mass and photon coupling 
$(m_a, g_{a\gamma\gamma})$ for the axion, and the 5 $\sigma$ 
discovery potential for the resonantly enhanced photon regeneration
experiment calculated for a configuration of 4 + 4 LHC dipole magnets.  
The existing exclusion limits indicated on the plot include the cavity
microwave experiments assuming axions saturate the dark matter halo
density \cite{Brad03}, the best direct solar axion search (CAST
collaboration) \cite{Ziou05}, the Horizontal Branch Star limit
\cite{Raff96}, and previous laser experiments \cite{Ruos92}.   
The red error ellipse indicates the positive result of the PVLAS 
collaboration, if interpreted as a light pseudoscalar, based on 
measurements of magnetically-induced dichroism of the vacuum
\cite{Zava06}.  For the estimated limits of resonantly enhanced 
photon regeneration presented here, the solid curve corresponds 
to the $\uparrow\uparrow\uparrow\uparrow$ configuration of the 
individual LHC dipole magnets in both the production and regeneration
strings; the dotted curve indicates the extension of the mass reach by 
additionally running in the $\uparrow\uparrow\downarrow\downarrow$, and
$\uparrow\downarrow\uparrow\downarrow$ configurations.}
\end{figure}

Defining a discovery to be 5$\sigma$, where $\sigma$ is
assumed to be dominated by the square-root counting statistics of the
dark-current background, we find that the experiment is sensitive to
axions or generalized pseudoscalars with 
$g_{a\gamma\gamma} = 2.7 \times 10^{-11}$ GeV$^{-1}$, 
after 10 days cumulative running, up to an
axion mass of  $m_a \sim 2 \times 10^{-4}$ eV (Figure 2).   Note the experiment's
reach in $g_{a\gamma\gamma}$ degrades sharply with increasing mass 
beyond that value.  This upper limit in mass for which one still has
effectively maximum sensitivity is dictated by the length of the 
dipole magnetic field $L$, as the momentum mismatch between a massless
photon and a massive axion  $q \sim L^{-1}$ defines the oscillation 
length of the problem.   As pointed out in ref. \cite{vanB87} however,
there is a practical strategy to extend the mass range upwards, if the
total magnetic length $L$ is comprised of a string of $N$ individual
identical dipoles of length $l$.  In this case, one may configure the
magnet string as a ``wiggler'' to cover higher regions of mass,
up to values corresponding to the oscillation length determined by a single
dipole, i.e. $q \sim l^{-1}$.  Figure 2 shows that the combination
of magnet configurations
$\uparrow\uparrow\uparrow\uparrow,~
\uparrow\uparrow\downarrow\downarrow,$ and 
$\uparrow\downarrow\uparrow\downarrow$ 
extend the mass reach up to $\sim 4 \times 10^{-4}$ eV.

While resonant photon
regeneration marks a significant improvement over the simple
experiment, in fact the sensitivity in $g_{a\gamma\gamma}$ still
only gains (or loses) as ${\cal F}^{1 / 2}$.  Thus in the
example above, the experiment would still reach a limit of
$g_{a\gamma\gamma} = 8.5 \times 10^{-11}$ GeV$^{-1}$ even if
the Fabry-Perot only achieved a finesse of ${\cal F} \simeq$ 30,000.
As it will likely be easier to attain higher values of ${\cal F}$ for shorter
baselines, we further note that as $g_{a\gamma\gamma} \propto (BL)^{-1}$,
a limit of $g_{a\gamma\gamma} \sim 10^{-10}$ GeV$^{-1}$,
i.e. equal to the CAST and the Horizontal Branch Star limit,
should still be achievable even with only {\it one} LHC dipole 
in each leg.

A fully-detailed derivation of the central result Eq. (12), and a detailed 
discussion of the experimental design worked out as an example in this Letter
will appear in a future publication.

We thank Frank Nezrick, Stan Whitcomb, Tom Carruthers, Kem Cook, 
Bruce Macintosh, Georg Raffelt and Guido Mueller for useful
conversations during the development of the concept.  This work was
supported in part by the U.S. Department of Energy under contracts
W-7405-ENG-48 and DE-FG02-97ER41029.  P.S. gratefully acknowledges 
the hospitality of the Aspen Center of Physics while working on this
project.

\end{document}